\documentclass[aps,twocolumn,superscriptaddress,noshowkeys,amssymb,amsmath,amsfonts,longbibliography]{revtex4-1}
\usepackage{bm}
\usepackage{graphicx}

\newcommand{\SSS}{\scriptscriptstyle}

\newcommand{\Ii}{{\rm i}}

\newcommand{\Ex}{\mathbf{e}_x}
\newcommand{\Ey}{\mathbf{e}_y}

\newcommand{\EpsM}{\epsilon_{{\SSS\text{M}}}}

\newcommand{\EpsD}{\epsilon_{{\SSS\text{D}}}}

\begin{document}

\title{Ultrafast fluorescent decay induced by metal-mediated dipole-dipole interaction in two-dimensional molecular aggregates}

\author{Qing Hu}
\affiliation{Department of Mechanical Engineering, Massachusetts Institute of Technology, Cambridge, MA 02139, USA}
\author{Dafei Jin}
\affiliation{Department of Mechanical Engineering, Massachusetts Institute of Technology, Cambridge, MA 02139, USA}
\author{Sang Hoon Nam}
\affiliation{Department of Mechanical Engineering, Massachusetts Institute of Technology, Cambridge, MA 02139, USA}
\author{Jun Xiao}
\affiliation{NSF Nanoscale Science and Engineering Center NSEC, University of California, Berkeley, CA 94720, USA}
\author{Yongmin Liu}
\affiliation{Department of Electrical and Computer Engineering, Northeastern University, Boston, MA 02115, USA}
\author{Xiang Zhang}
\affiliation{NSF Nanoscale Science and Engineering Center NSEC, University of California, Berkeley, CA 94720, USA}
\affiliation{Material Sciences Division, Lawrence Berkeley National Laboratory, Berkeley, CA 94720, USA}
\affiliation{Department of Physics, King Abdulaziz University, Jeddah 21589, Saudi Arabia}
\author{Nicholas X. Fang}\email{nicfang@mit.edu}
\affiliation{Department of Mechanical Engineering, Massachusetts Institute of Technology, Cambridge, MA 02139, USA}

\date{\today}
\maketitle
\pretolerance=8000 % good to put here,otherwise may change the affiliation style

\textbf{Two-dimensional molecular aggregate (2DMA), a thin sheet of strongly interacting dipole molecules self-assembled at close distance on an ordered lattice, is a fascinating fluorescent material. It is distinctively different from the single or colloidal dye molecules or quantum dots in most previous research. In this paper, we verify for the first time that when a 2DMA is placed at a nanometric distance from a metallic substrate, the strong and coherent interaction between the dipoles inside the 2DMA dominates its fluorescent decay at picosecond timescale. Our streak-camera lifetime measurement and interacting lattice-dipole calculation reveal that the metal-mediated dipole-dipole interaction shortens the fluorescent lifetime to about one half and increases the energy dissipation rate by ten times than expected from the noninteracting single-dipole picture. Our finding can enrich our understanding of nanoscale energy transfer in molecular excitonic systems and may designate a new direction for developing fast and efficient optoelectronic devices.}

How a fluorescent nano-emitter releases its energy to the environment is a longstanding research topic in nanoscale light-energy collection and conversion \cite{Andrew2000Science,Herek2002Nature,Coles2014NatMater,Rabouw2014NatCommun}. In the past decades, there have been considerable investigations on the fluorescence enhancement and quenching of a nano-emitter influenced by a structured environment \cite{Dulkeith2002PRL,Anger2006PRL,Kinkhabwala2009NatPhoton,Woolf2014PNAS}. It is known that in a lossless medium, the emitter can decay radiatively by emitting photons or nonradiatively by generating molecular vibrations \cite{ChanceBook,Novotny1996APL,Kasha1963RadRes}. The typical fluorescence lifetime due to these two dissipation channels are of the order of nanoseconds \cite{AmerongenBook,Gersten1981JCP}. By contrast, in the proximity of a lossy medium such as a metallic substrate, the emitter can also decay nonradiatively through transferring energy into collective electron oscillations. The strength of this dissipation channel usually dominates over the above two channels and results in a significantly shortened fluorescence lifetime down to tens of picoseconds \cite{Chance1974JCP,Lakowicz2005}.

So far, most of the nano-emitters studied are single dye molecule (DM), single quantum dot (QD), or colloids of randomly dispersed DMs or QDs \cite{Dulkeith2002PRL,Anger2006PRL,Moreau2012Nature}. In such systems, each emitter can be well described by a single dipole which interacts exclusively with the environment. Owing to the sparsity and randomness of the dipole distribution, interaction between different dipoles at different location is considered negligible in earlier work \cite{Stuart1998PRL}. Strikingly, however, the so-called molecular aggregate (MA) belongs to a unique class of nano-emitters that behave rather distinctively from the above. Each MA can be envisioned as a collection of self-assembled dipoles arranged at a close distance ($\sim1$~nm) on an ordered molecular lattice \cite{KobayashiBook,Saikin2013Nanophotonics,Arias2013PhysChem}. The dipoles situated on the different lattice sites interact coherently and strongly with each other through the dipole-dipole interaction. This interaction causes the energy levels to recombine and form  blue-shifted H-band or red-shifted J-band, producing the so-called H-aggregate or J-aggregate respectively \cite{Eisfeld2006ChemPhy,Spano2009AccChemRes}. The fascinating MA can exhibit strong exciton-photon coupling \cite{Lidzey1998Nature,Zheng2010AM,Fofang2011NanoLett} and superradiance \cite{Meinardi2003PRL,Fidder1990CPL}, and have been used to demonstrate many fundamental phenomena, such as Rabi splitting \cite{Lidzey1999PRL,Bellessa2004PRL,Vasa2013NatPhoton} and room-temperature Bose-Einstein condensation \cite{Klaers2010Nature,Plumhof2013NatMater}, and have been applied to design various devices, such as organic light-emitting diodes \cite{Tischler2005PRL}, solar cells \cite{WurfelBook} and light-harvesting organic antennas \cite{Savolainen2008PNAS,Hildner2013Science}.

In this work, we identify for the first time the dominant role of dipole-dipole interaction inside a two-dimensional molecular aggregate (2DMA), in determining its ultrafast quenching and energy-dissipation rate. Through a series of delicate molecule-level fabrication and picosecond-timescale measurement, we have observed a greatly shortened fluorescence lifetime down to only 5~ps or less. This observation can only be interpreted by our interacting lattice-dipole model, as opposed to the conventional single-dipole model where the internal interaction is completely ignored. Based on the interacting-dipole model, we are able to demonstrate that the amplitude and phase of this interaction between every pair of dipoles is strongly affected by a metallic substrate at a precisely controlled nanometric distance from the 2DMA. The metal-mediated dipole-dipole interaction inside the 2DMA leads to at least ten times greater energy dissipation rate than that commonly expected from the single-dipole picture. Our finding can enrich our understanding to nanoscale energy transfer in molecular excitonic systems, and can provide useful insight into many other 2D excitonic materials that are attracting intense research interest in the recent years \cite{Amani2015Science,Sie2015NatMater,Akselrod2015NanoLett}.

\begin{figure}[htb]
\centering
\includegraphics[scale=0.35]{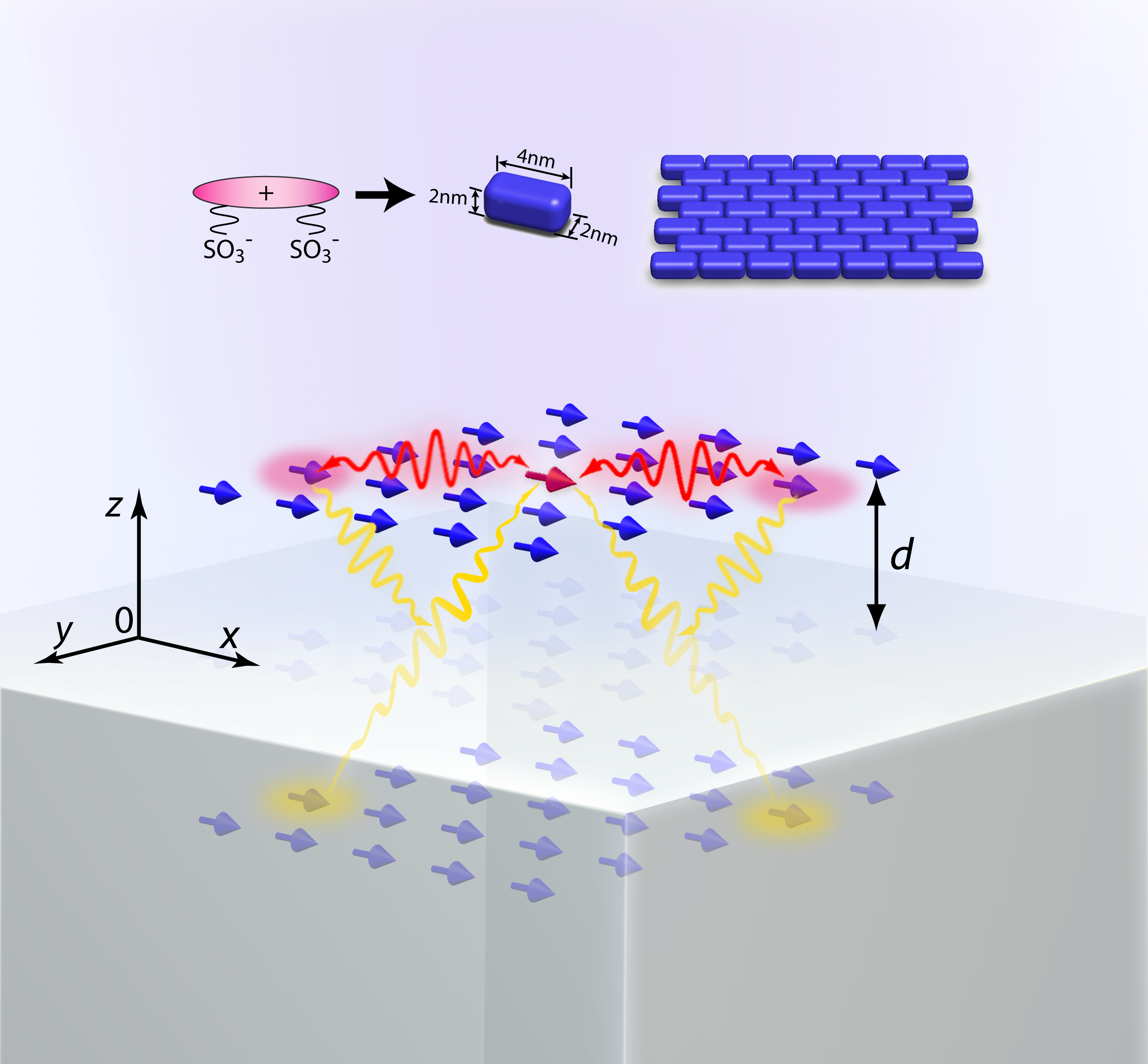}
\caption{Schematics of the model system and the fabricated sample structure. The 2D molecular aggregate is placed above the metal surface at a distance $d$. The aggregate has a brickstone lattice. The molecules (dipoles) are all oriented along the same $x$ direction. The interactions lie between real and image dipoles.}
\label{fig:Fig1}
\end{figure}

To begin with, we conceptually illustrate our system in Fig.~1. A 2D dipole array representing a monolayer J-aggregate is positioned near a silver substrate at a distance $d$. Due to self-assembly, a group of dipoles (of a number $N$) automatically form a brickstone lattice (see Materials and Methods) \cite{KobayashiBook}, and are all polarized along the long axis of the molecules (which we set as $x$-direction). The dynamics of this interacting-dipole system must be described by a set of coupled equations \cite{Philpott1975PRB},
\begin{equation}
\sum_{s'} \left( \mathbf{1} \delta_{ss'} + \bm{\alpha}\cdot\bm{G}_{ss'} \right) \cdot \bm{p}_{s'}  = \bm{\alpha}\cdot\bm{E}_{\text{inc}},\ (s=1,2,\dots,N),\label{EqLatticDipoleModel}
\end{equation}
in which $s$ and $s'$ label the sites where the dipoles situate. $\bm{\alpha}=\alpha\Ex\Ex$ is the polarizability of monomer, which contains a monomer resonance frequency $\omega_0$ (see the detailed form below). $\bm{p}=p\Ex$ is the dipole moment that relates to $\bm{\alpha}$ under a total electric field $\bm{E}$ by $\bm{p} = \bm{\alpha}\cdot\bm{E}$.  $\bm{E}_{\text{inc}}$ is an incident electric field and $\bm{G}_{ss'}$ is the interaction tensor. In this system, a nonretarded Coulombic form of $\bm{G}_{ss'}$ suffices \cite{NovotnyBook,JacksonBook},
\begin{align}
\bm{G}_{ss'} =& \frac{1}{4\pi\EpsD} \left\{ \left(\frac{\Ex\Ex}{R_{ss'}^3} -\frac{3\bm{R}_{ss'}\bm{R}_{ss'}}{R_{ss'}^5}\right) (1-\delta_{ss'}) \right. \nonumber \\
& + \eta \left. \left(\frac{\Ex\Ex}{Q_{ss'}^3} -\frac{3\bm{R}_{ss'}\bm{R}_{ss'}}{Q_{ss'}^5}\right) \right\},
\label{EqInteractionTensor}
\end{align}
where an important coefficient $\eta$ due to the presence of substrate reads
\begin{equation}
\eta=\frac{\EpsD-\EpsM}{\EpsD+\EpsM}.
\end{equation}
Here $\EpsD$ and $\EpsM$ are the permittivities of dielectric and metal in the upper and lower half space, respectively.
The first term of \eqref{EqInteractionTensor} represents the interaction between two real dipoles inside the dielectric; the factor $1-\delta_{ss'}$ removes the unphysical self-interaction. $\bm{R}_{ss'}=(x_s-x_{s'})\Ex+(y_s-y_{s'})\Ey$ is the displacement vector between the two real dipoles $s$ and $s'$; $R_{ss'}=\sqrt{(x_s-x_{s'})^2+(y_s-y_{s'})^2}$ is its magnitude. The second term represents the interaction between a real dipole in the dielectric and an image dipole in the metal. The coefficient $\eta$ determines the amplitude and phase of an image dipole relative to its corresponding real dipole \cite{NovotnyBook}.  $Q_{ss'}=\sqrt{(x_s-x_{s'})^2+(y_s-y_{s'})^2+(2d)^2}$ takes account of an additional distance $2d$ between a real dipole and an image dipole. As is known, the first term in $\bm{G}_{ss'}$ gives rise to the formation of red-shifted J-band from the monomer resonance frequency. It does not include any effect from a dissipative substrate. On the contrary, the second term takes account of the dissipative motion of electrons in the metal through $\EpsM$. Once the metal is placed within a distance of several nanometers to the 2DMA, this term dominates the nonradiative decay. In this sense, the metal mediation to the dipole-dipole interaction acts upon every pair of dipoles inside the 2DMA, accompanying the J-band formation.

\begin{figure*}[htb]
\centering
\includegraphics[scale=0.45]{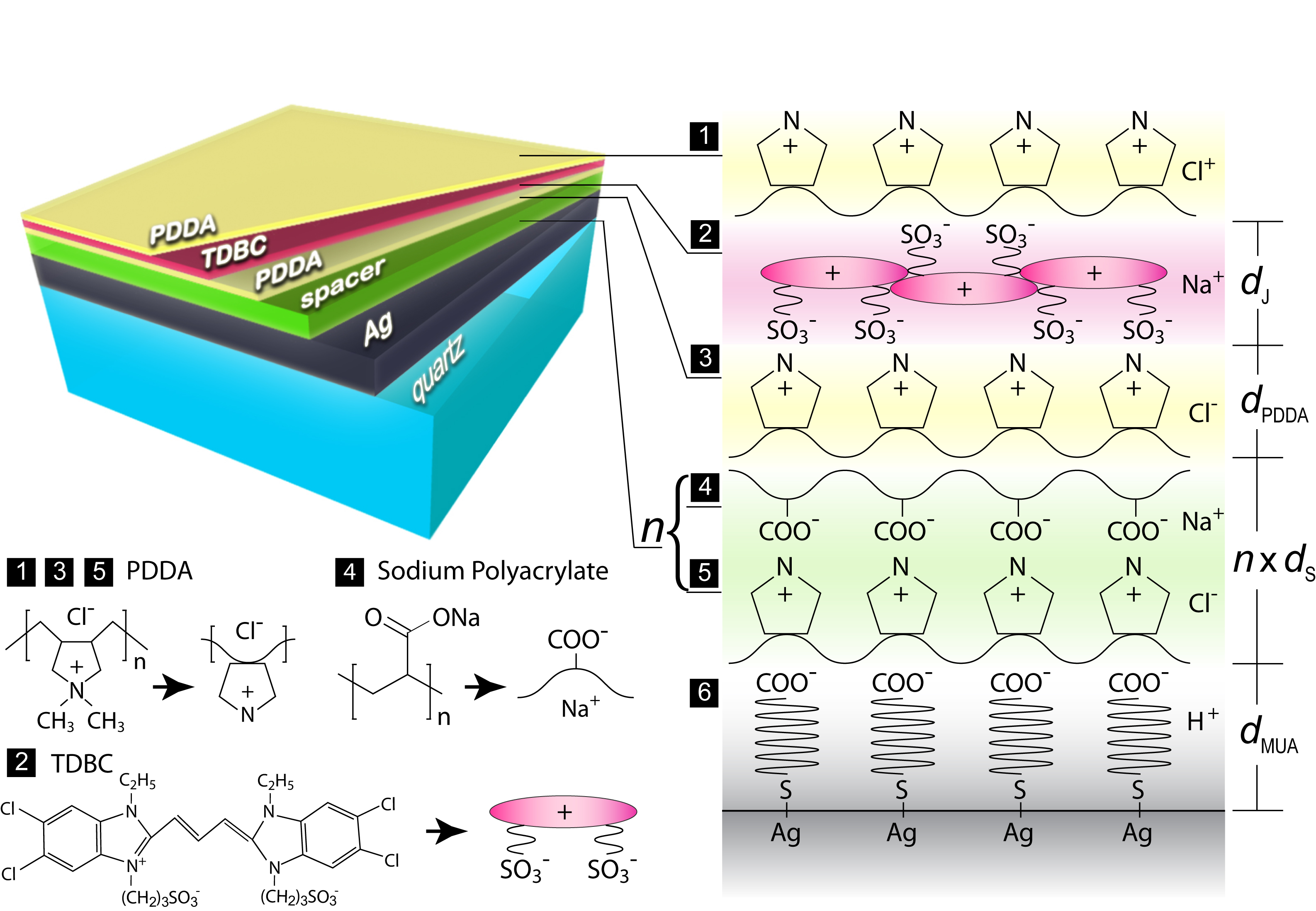}
\caption{Schematics of the fabricated molecular multilayer structure. The layers are formed one by one via the adsorption between the positively and negative charged molecules. From the bottom to the top, it consists of 100~nm Ag, MUA bonding layer, (PDDA/PolyArc)$_n$ spacer layer, and (PDDA/TDBC/PDDA) cyanine layer encapsulated in PDDA.}
\label{fig:Fig2}
\end{figure*}

As a comparison, we may attempt to follow the conventional treatment and consider the entire 2DMA as a single dipole of an effective polarizability $\bm{\alpha}_{\text{eff}}$ resonant at the free-space J-band frequency $\omega_{\SSS\text{J}}$ \cite{Ellenbogen2011PRB}. Here the effective $\bm{\alpha}_{\text{eff}}$ and the monomer $\bm{\alpha}$ are mutually related by $\bm{\alpha}_{\text{eff}}=C\bm{\alpha}|_{\omega_0\rightarrow\omega_{\SSS\text{J}}}$, where $C$ is a ``normalization" constant that can be determined from calculation. Then the dynamic equation becomes
\begin{equation}
\bm{p}_{\text{eff}}+\bm{\alpha}_{\text{eff}}\cdot\bm{G}\cdot\bm{p}_{\text{eff}} = \bm{\alpha}_{\text{eff}} \cdot\bm{E}_{\text{inc}},\label{EqSingleDipoleModel}
\end{equation}
where $\bm{p}_{\text{eff}}$ is the effective single-dipole moment, and $\bm{G}$ only takes care of the interaction between the effective single dipole as a whole and its image dipole,
\begin{equation}
\bm{G} = \frac{\eta}{8d^3} \Ex\Ex.\label{EqSingleDipoleInteraction}
\end{equation}
In this picture, the internal dipole-dipole interaction inside the 2DMA and mediated by the metallic substrate has been completely ignored. As shown below, because of this missing nonradiative decay channel encapsulated in the metal-mediated dipole-dipole interaction, the conventional picture leads to a large discrepancy between the theoretical prediction and experimental result.

\section*{Materials and Methods}

\noindent\textbf{Molecular multilayer growth.}

In order to accurately characterize the complex system of 2DMA on a metallic substrate, we need to choose a representative 2DMA which has a stable structure and a strong radiation power. It is known that the J-aggregate of dye 5,5',6,6'-tetrachloro-1,1'-diethy1-3,3'-di(4-sulfobuty1)-benzimidazolocarbocyanine (TDBC) is a superior molecular aggregate with coherently coupled transition dipoles. It shows a very intense, narrow, and redshifted J-band with respect to the monomer band, and is chemically compatible with various nanofabrication techniques. It has been considered as an ideal mesoscopic system bridging single molecules and large size crystals, and has been considered as a promising material for excitonic devices \cite{Tischler2007OrgElectron}.

Our experiment requires deposition of well-ordered molecular monolayer on high-quality metal surface and a precise control of the spacer thickness. We achieve this by molecular layer-by-layer (LBL) adsorption \cite{Fukumotoa1998TSF,Tischler2007OrgElectron} (schematically shown in \ref{fig:Fig2}). This method has been proved to be reliable, reproducible, and easily operated \cite{Stockton1997Macromolecules,Decher1997Science,Ariga1997JACS,DecherBook}. It has been employed in designing molecular devices in the past decades \cite{Tischler2005PRL,Tischler2007OrgElectron}. The substrate is prepared by evaporating 100~nm thick Ag on a well cleaned super flat ($<$5~{\AA} roughness) quartz plate. The evaporation is run at high vacuum (1E-6~Torr) and slow deposition rate (0.8~{\AA}/s) to guarantee a smooth surface. The fresh-made Ag substrate is then immersed in a 0.1~M 11-Mercaptoundecanoic (11-MUA) acid aqueous solution for a conformal coating of a monolayer of decanoic acid. The Ag with the decanoic acid group on top carries negative charges and is able to adsorb PDDA (a cationic polyelectrolyte). The anionic polyelectrolyte sodium polyacrylate (PolyArc) is then adsorbed onto the positively charged surface of PDDA layer. The sequential alternating adsorption of PDDA and PolyArc yields $n$ layers of (PDDA/PolyArc) assembly, denoted as (PDDA/PolyArc)$_n$. It provides a spacer of precisely controlled thickness $n\times d_s$ at molecular level (shown in green color on the top left of Fig.~\ref{fig:Fig2} and labeled by ``4" and ``5"). Because the anionic cyanine molecules TDBC can adsorb a positively charged PDDA layer to form a J-aggregate monolayer, the sequential alternate adsorption of PDDA and TDBC yields another assembly of (PDDA/TDBC/PDDA) on top of the spacer with the thickness $d_{\SSS\text{J}}$. Briefly, the molecular multilayer structure can be denoted as MUA-(PDDA/PolyArc)$_n$-(PDDA/TDBC/PDDA) from the bottom to the top.

\begin{figure*}[htb]
\centering
\includegraphics[scale=0.56]{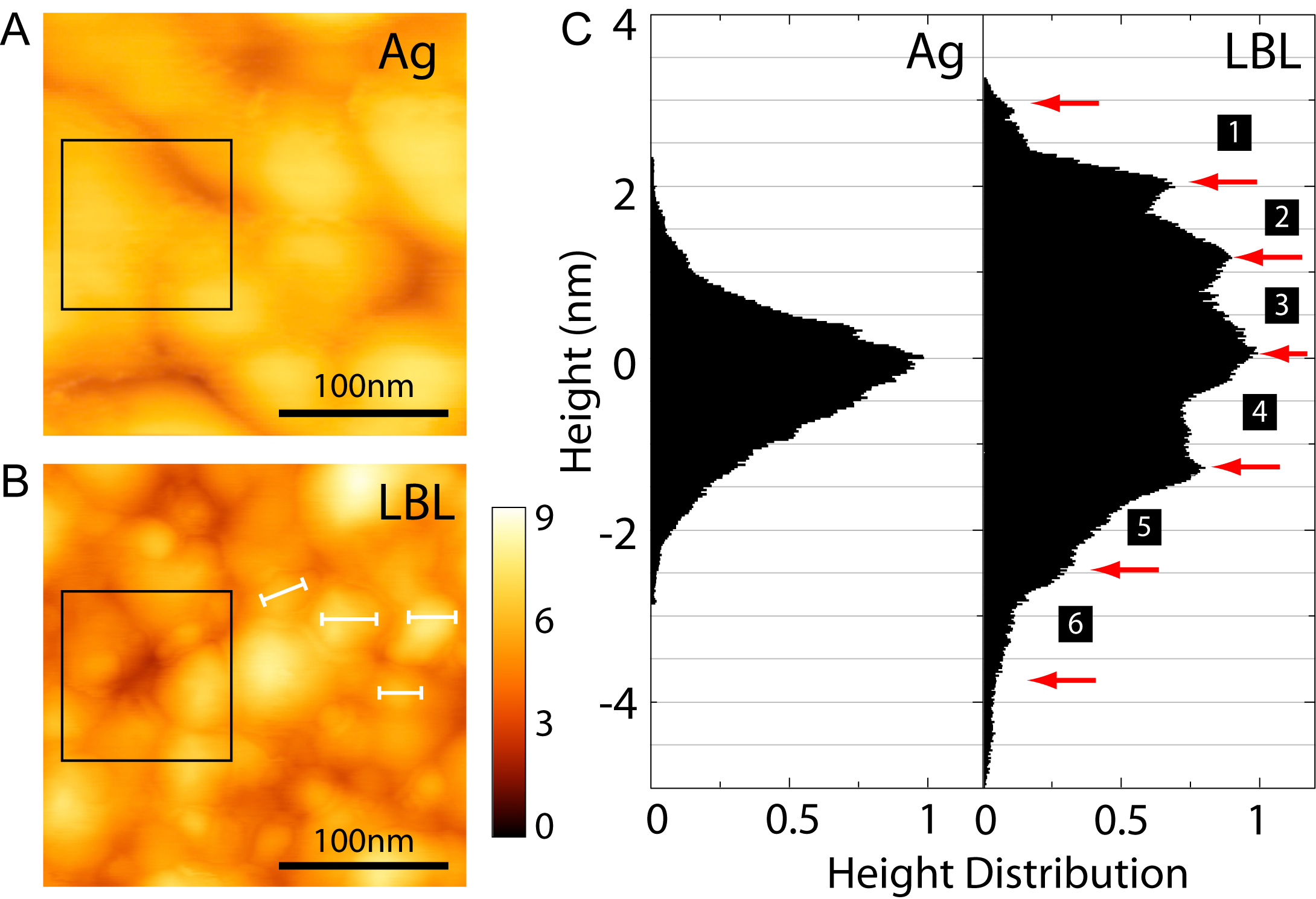}
\caption{Morphologic characterization of the layer-by-layer structure of MUA-(PDDA/PolyArc)-(PDDA/TDBC/PDDA). (A) Scanned AFM image for bare 100~nm Ag on quartz. (B) Scanned AFM image for molecular multilayer on Ag. The size of the scanning area is 0.25~$\mu$m by 0.25~$\mu$m. (C) The statistic height distribution of the sample areas marked by the black boxes in (A) and (B). The red arrows indicate the discrete steps labeled with numbers. Each number corresponds to a layer labeled in Fig.~\ref{fig:Fig2}.}
\label{fig:Fig3}
\end{figure*}

\noindent\textbf{Morphology characterization.}

The LBL thin films grown on Ag surface possess remarkable morphology as layered materials bearing nanometric thickness variation \cite{Tischler2007OrgElectron}. The topography can be verified by atomic force microscopy (AFM), which also provides a way to determine the thickness of each molecular layer. Figure~\ref{fig:Fig3} shows the AFM (Cypher\textsuperscript{\sffamily\tiny TM}AFM) scanned surface morphology of the samples with and without coated molecules. The 100~nm Ag by electron-beam evaporation on quartz has an around 100~nm grain size on average (Fig.~\ref{fig:Fig3}A). The square root roughness is around 1.98~nm. After completing the molecule deposition as shown in Fig.~\ref{fig:Fig2}, the surface morphology changes. In Fig.~\ref{fig:Fig3}B, we can observe the lumped regions with clear boundaries. They are the evidence of J-aggregate domains \cite{KobayashiBook,Saikin2013Nanophotonics}. The domain size (labeled with white lines) is about 20~nm and the square root roughness becomes 4.16~nm. It is known that each TDBC monomer has the in-plane dimension from 1~nm to 3~nm \cite{Valleau2012JCP}, so the total number of the molecules in one domain is around 50, which is consistent with the literature \cite{Moll1995JCP}. Based on the morphology data, a histogram of height distribution can be obtained by the statistics of height over the sample area. We randomly select a 100~nm-by-100~nm area on a bare-Ag sample (shown in the black box in Fig.~\ref{fig:Fig3}A). The histogram (marked with ``Ag" in Fig.~\ref{fig:Fig3}C) shows a concentrated distribution at the center and the total variation is over 8~nm. To see the multilayered feature of the LBL sample, we carefully select an area which contains defects and holes inside the molecular layers. The statistical histogram over this area exhibits discrete steps (marked with numbers in Fig.~\ref{fig:Fig3}C) \cite{Tischler2007OrgElectron}. The jump between every two peaks corresponds to the thickness of one molecular layer. Therefore, the distribution demonstrates the molecular multilayer structure under the LBL process. According to this diagram we can obtain the thickness of each monolayer from the top to the bottom to be around 1.0~nm, 0.9~nm, 1.2~nm, 1.3~nm, 1.2~nm, 1.5~nm, respectively. Hence we know that one spacer layer is around 2.5~nm thick ($d_\text{s}$), the thickness of 11-MUA layer ($d_{\SSS\text{MUA}}$) is around 1.5~nm, and the PDDA ($d_{\SSS\text{PDDA}}$) is around 1.2~nm. To systematically study the effect of distance $d$ on the fluorescent behavior, we fabricate a series of samples consisting of one pair of PDDA and TDBC layer, but different numbers of spacer layers; i.e., $n$ varies from 1 to 6. This provides a fine-tuned change of $d=d_{\SSS\text{MUA}}+n\times{d_\text{s}}+d_{\SSS\text{PDDA}}$ from 5.0~nm to 16.5~nm.

\begin{figure*}[htb]
\centering
\includegraphics[scale=0.56]{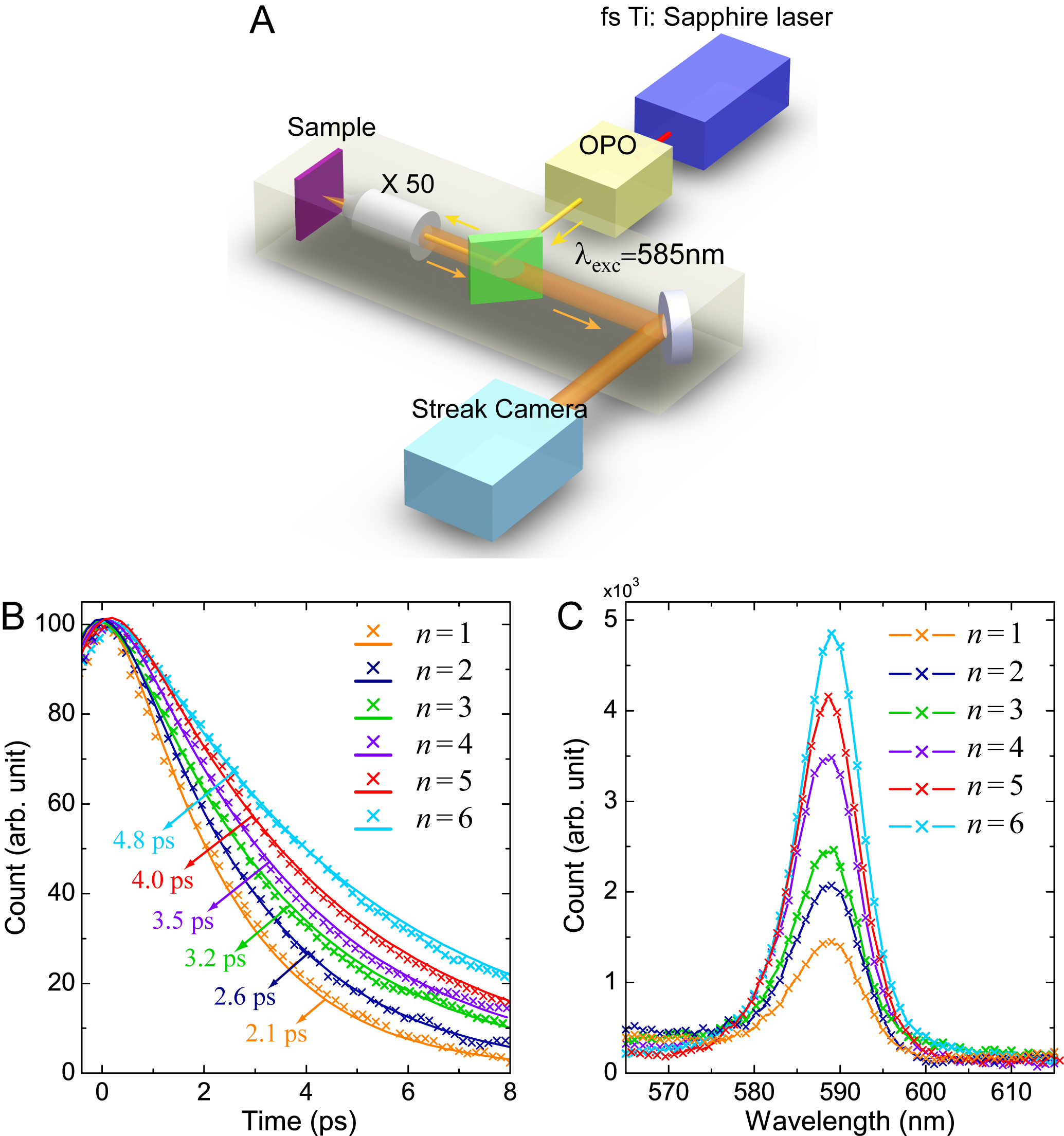}
\caption{Lifetime and photoluminescence measurement for the molecular multilayer structures of MUA-(PDDA/PolyArc)$_n$-(PDDA/TDBC/PDDA) with $n = 1, 2, \dots , 6$. (A) Schematic of experimental setup. (B) Time-resolved photon counts (marks) and  exponential fitting curves (solid lines). (C) Photoluminescence intensity. The structures with different numbers of spacer layers from 1 to 6 are shown in different colors.}
\label{fig:Fig4}
\end{figure*}

\begin{figure*}[htb]
\centering
\includegraphics[scale=0.56]{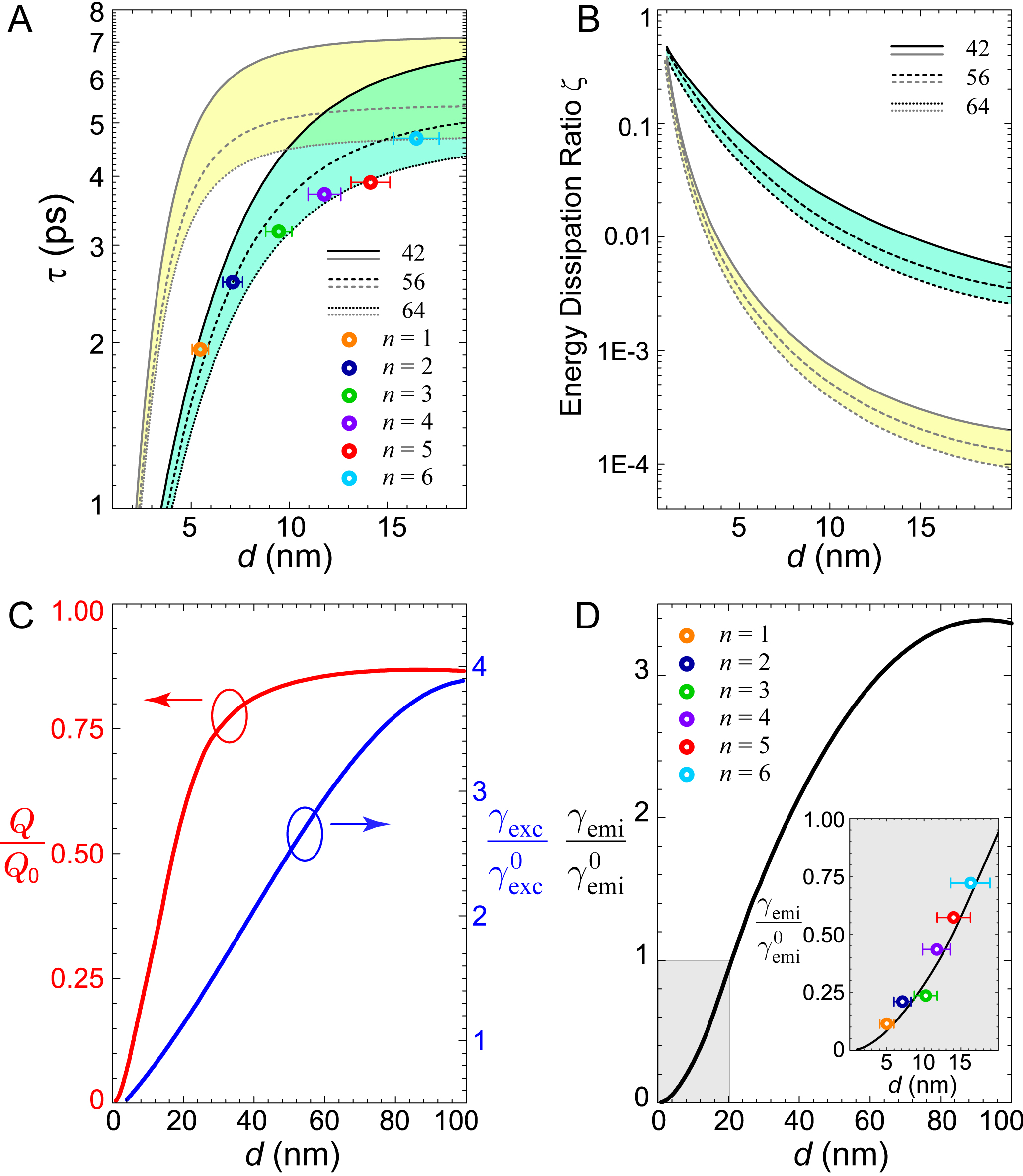}
\caption{Theoretical analysis of the fluorescent behavior of the J-aggregate on Ag substrate with different numbers of spacer layers. Calculated lifetime (A) and energy dissipation ratio (B) based on the single-dipole model (yellow region) and lattice-dipole model (green region) as a function of distance $d$ from the substrate. In the calculation, the total number of dipoles in the array is set to be 42, 56, and 64 for both models, and is shown in the solid, dashed and dot lines, respectively. The measured lifetime from Fig.~\ref{fig:Fig4}(A) is plotted with open circles in Fig.~\ref{fig:Fig5}(A). Only the lattice-dipole model agrees with the measurement. (C) The calculated normalized excitation rate (in red) and normalized radiation rate (in black) as a function of $d$. (D) Calculated normalized emission rate as a function of $d$. The inset shows the enlarged near-surface region ($d<20$~nm). The measured result from Fig.~\ref{fig:Fig4}(B) is plotted with open circles. For the plots of the measurement, the x-error bars indicate the uncertainty in $d$ due to the mean roughness of deposited molecular films.}
\label{fig:Fig5}
\end{figure*}

\section*{Results and Discussion}

\noindent\textbf{Fluorescence lifetime and photoluminescence measurement.}

The fluorescence lifetime of J-aggregate is measured by a picosecond streak camera setup as shown in Fig.~\ref{fig:Fig4}A. The excitation light is carried out with an optical parametric oscillator (Spectra Physics, Inspire HF~100) pumped by a mode-locked Ti:sapphire oscillator. The laser pulse width is about 200~fs and the repetition rate is 80~MHz. The light is guided into a Zeiss inverted microscope (Axiovert 200) and focused onto the sample by a Zeiss 50x objective. The emission signal is detected in the reflection configuration, and the signal passing through a bandpass filter with a bandwidth of 30~meV is collected by a synchroscan Hamamatsu streak camera (C10910-02), whose overall time resolution is 2~ps. The transmissivity of the optical system is carefully calibrated to evaluate the absolute power level at the focusing plane. The laser pulse width is measured by a home-built autocorrelator at the focus throughout the scanning range. The incident wavelength is chosen at 585~nm for on-resonance excitation of TDBC. For each sample, the measurement is repeated several times in different areas of samples and the final result is the average of multiple measurements. In Fig.~\ref{fig:Fig4}B, the measured time-resolved fluorescent intensity of the molecular structures with 1 to 6 spacer layers are shown by the marks of different colors. The exponential fitting (solid lines in the same color) assisted with a convolution algorithm \cite{LakowiczBook,Hong2014NatNanotech} indicates the increase of lifetime with the increasing distance $d$. The overall lifetime is only several picoseconds, extremely short compared with the situation of a single molecule on silver, which is around ten picoseconds or larger \cite{Lakowicz2005,Cnossen1993JCP,Vaubel1971CPL}.

We have also measured the photoluminescence spectrum (Horiba Fluorolog-3) of our molecular structures of different $d$ as shown in Fig.~\ref{fig:Fig4}C. The measured data exhibit the trend that the fluorescence power increases with the increasing thickness of spacer. This agrees with the expectation of a reduced nonradiative decay when the fluorescence material is brought away from the lossy medium.

\noindent\textbf{Theoretical analysis and comparison to experiment.}

Our theoretical analysis for the fluorescent behaviors is based on our interacting lattice-dipole model. Each TDBC monomer can be treated as a two-level oscillator with the Lorentzian polarizability \cite{Fukumotoa1998TSF},
\begin{equation}
4\pi \bm{\alpha}(\omega) = 4\pi\alpha(\omega)\Ex\Ex =\frac{\omega_0^2f_0}{\omega_0^2-\omega^2-\Ii\omega\gamma_0}\Ex\Ex,
\label{EqPolaribility}
\end{equation}
where $f_0$ is the oscillator strength, $\gamma_0$ gives a free-space fluorescence lifetime about 4~ns \cite{Kemnitz1990JCP}, and $\omega_0$ gives a 520~nm monomer resonance wavelength. With a brickstone pattern and an adjustable $f_0$, the resonance wavelength for the J-aggregate can be tuned to be around 590~nm, which is the J-band of TDBC aggregate \cite{KobayashiBook,Saikin2013Nanophotonics}. In the absence of an incident field, we can summarize ~\eqref{EqLatticDipoleModel} and \eqref{EqInteractionTensor} into a matrix eigenvalue problem,
\begin{equation}
\left(\omega^2+\Ii\omega\gamma_0-\omega_0^2\right)\mathbf{G} = \mathbf{G}\cdot\mathbf{P},
\label{EqEigenProblem}
\end{equation}
where $\mathbf{P}$ is a column vector $(p_1,p_2,...p_{\SSS N})^{\SSS\text{T}}$ and $\mathbf{G}$ is a N-by-N matrix determined by \eqref{EqInteractionTensor}. The eigenfrequencies solved from \eqref{EqEigenProblem} are generally complex-valued, $\tilde{\omega}=\omega_{\SSS\text{J}}-\frac{\Ii}{2}\gamma$. The imaginary part gives the lifetime of the mode $\tau=\gamma^{-1}$ from the nonradiative decay. The lattice-dipole mode profile can be obtained by the eigenstates of $\mathbf{P}$. The only relevant mode under far-field light illumination is the so-called bright mode, which has the least phase variation and can be picked out numerically.

The calculated lifetime of bright mode as a function of the distance $d$ is plotted in Fig.~\ref{fig:Fig5}A. The green color shows a region with variable total dipole number $N$. The colored open circles are from the measurement (shown in Fig.~\ref{fig:Fig4}A). One can see that our theoretical calculation matches the trend of the experimental measurement very well. For comparison, we have also calculated the lifetime using the conventional single-dipole model (refer to \eqref{EqSingleDipoleModel} and \eqref{EqSingleDipoleInteraction}). It is apparent that the trend (shown in yellow) deviates significantly from the measurement. Therefore, the actual lifetime of the J-aggregate is much shorter than the expectation from the single-dipole picture. It reveals that a major dissipation channel comes from the nonradiative decay via electron oscillations inside the metal in every step of dipole-dipole interaction.

With our solved eigenfrequencies and eigenstates, we are able to analyze the energy dissipation and fluorescent properties in a greater detail. At a close distance to the metal surface, the emissive energy of the 2DMA is mostly absorbed by the electron collision inside the metal. We can define an energy dissipation ratio as the energy flux towards the metal versus the total energy flux in all directions, $\zeta=\frac{\text{Re}[\int_{\sigma^-}\bm{S}\cdot\text{d}\bm\sigma]}{\text{Re}[\int_{\sigma}\bm{S}\cdot\text{d}\bm\sigma]}$, where $\bm{S}$ is the energy flux density calculated using the retarded formula of dipole radiation \cite{JacksonBook}, $\sigma$ is a closed surface surrounding the dipole lattice and $\sigma^-$ is the integration area beneath the 2DMA but above the substrate. Fig.~\ref{fig:Fig5}B shows $\zeta$ as a function of $d$ using our lattice-dipole model in comparison with the conventional single-dipole model. Clearly, the energy dissipation rate in the lattice-dipole picture is about ten times greater than that in the single-dipole picture. Thus in reality much more emissive energy is transferred to the lossy metal substrate than commonly expected.

\begin{figure*}[htb]
\centering
\includegraphics[scale=0.56]{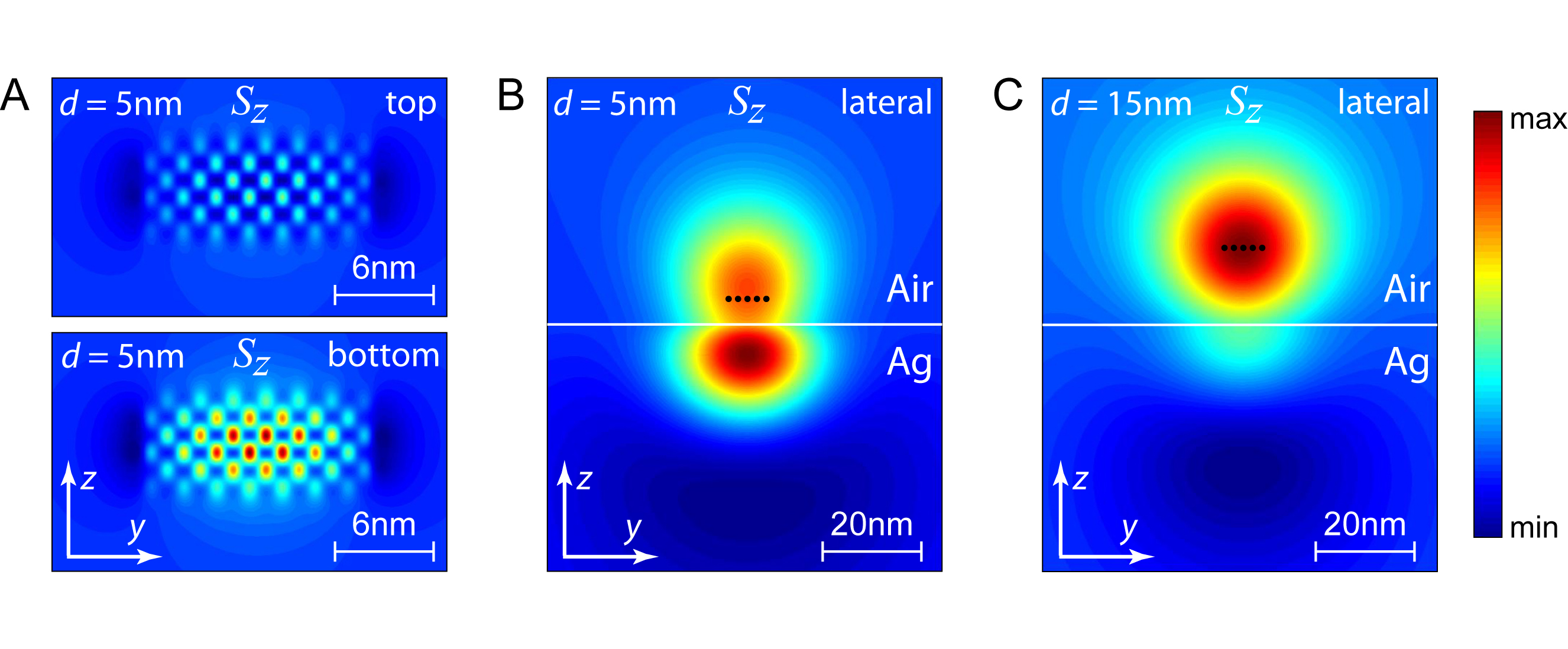}
\caption{Distribution of the energy flux density of the system. (A) Top view and bottom view of the distribution on the dipole lattice. The viewing plane ($xy$) is located at 5~nm above (labeled as ``top") and below (labeled as ``bottom") from the dipole plane. Here $d$ is chosen as 5~nm. (B) and (C) Lateral views of the distribution on the dipole lattice. The viewing plane ($yz$) is located 30~nm away from the edge of the dipole lattice. $d$ is 5~nm for (B) and 15~nm for (C).}
\label{fig:Fig6}
\end{figure*}

The fluorescence process involves excitation and radiation \cite{Anger2006PRL,Kinkhabwala2009NatPhoton,Dulkeith2002PRL}. The abilities of excitation and radiation can be represented by the emission rate $\gamma_\text{emi}$, and the quantum yield $Q$ defined as $Q=\frac{\gamma_\text{rad}}{\gamma}$, where $\gamma_\text{rad}$ and $\gamma$ are the radiative decay rate and total decay rate respectively. So the emission rate can be written as $\gamma_\text{emi}=\gamma_\text{exc}Q=\gamma_\text{exc}(\frac{\gamma_\text{rad}}{\gamma})$. The dipoles above a metallic substrate are excited by the addition of incident field $\bm{E}_\text{inc}$ and reflected field from the substrate $r\bm{E}_\text{inc}$. The normalized excitation rate can be expressed as
\begin{equation}
\frac{\gamma_\text{exc}}{\gamma^0_\text{exc}}={\left|\frac{\Ex\cdot[\bm{E}_\text{inc}+r\bm{E}_\text{inc}]}{\Ex\cdot[\bm{E}_\text{inc}+r_0\bm{E}_\text{inc}]} \right|}^2
\label{EqExcitationRate}
\end{equation}
where $r$ and $r_0$ are the reflection coefficients of the silver substrate and a reference quartz substrate, respectively. The quantum yield can be calculated by the ratio of time-averaged energy flux towards the upper space versus the total energy flux towards the whole space, and so can be expressed as
\begin{equation}
Q=\frac{\gamma_\text{rad}}{\gamma}=\frac{\text{Re}[\int_{\sigma^+}\bm{S}\cdot\text{d}\bm\sigma]}{\text{Re}[\int_{\sigma}\bm{S}\cdot\text{d}\bm\sigma]},
\label{EqQuantumYield}
\end{equation}
where $\sigma^+$ is the part of $\sigma$ facing the upper half-space. In order to normalize $Q$, we choose a bare quartz without Ag as the substrate to calculate $Q_0$.

Fig.~\ref{fig:Fig5}C shows the normalized quantum yield $Q/Q_0$ and excitation rate as a function of $d$. One can see that the normalized quantum yield $Q$ (plotted in red) increases with increasing $d$. It reaches a saturation value when $d$ approaches 50~nm. Fig.~\ref{fig:Fig5}D shows the normalized emission rate of the 2DMA. It is strongly suppressed as $d$ gets smaller than 20~nm. The enlarged part of the curve for this short distance (in the shaded region) is shown in the inset, where the measured results from Fig.~\ref{fig:Fig4}C are labeled by open circles. Note the measurement shown here is also normalized to the corresponding quartz samples. Our theoretical calculation matches the measurement very well. This further proves that the interacting lattice-dipole model successfully describes the 2DMA near a metallic substrate.

As an example, we also calculate the energy flux density distribution of an 8-by-7 dipole lattice. Fig.~\ref{fig:Fig6}A shows the distribution of the amplitude of the forward and backward energy flux density component $S_z$ relative to the substrate surface viewed from the top and bottom, respectively. One can clearly see that the backward energy flux density is much stronger than the forward one. The backward energy flux eventually dissipates into the metal. In addition, the color map clearly exhibits a brickstone lattice pattern, from which we can see that the mode amplitude is stronger in the center and weaker on the edge, and the whole lattice exhibits a collective oscillation. It displays the fine structures inside the dipole lattice; namely, the lattice cannot be simply envisioned as an effective single dipole. The lateral distribution of the energy flux $S_z$ (Fig.~\ref{fig:Fig6}B) shows that more energy is confined inside the metal region, giving a visualization to the intense energy dissipation induced by the metallic substrate. As a comparison, we plot the $S_z$ of the dipole lattice at a greater distance $d=15$~nm in Fig.~\ref{fig:Fig6}C. In that case, the energy dissipated into the metal is reduced.

In conclusion, we have investigated the fluorescent behaviors of 2D molecular aggregates at different distances from a metallic substrate, by measuring the lifetime and photoluminescence intensity. Our study shows that when the molecular aggregates are close to the metal surface, the dipole-dipole interaction mediated by the metal plays a dominant role in the nonradiative decay. It leads to an ultrafast fluorescent decay and ultrastrong energy dissipation. The studies of coupled systems of molecular aggregate and metal can deepen our understanding of the interaction between nano-emitters and nanostructures \cite{Alu2009PRL,Mueller2013PRB}. Our findings provide new guidelines to design and optimize fast and efficient molecular optoelectronic devices \cite{Walker2011NanoLett}.

\section*{Acknowledgements}

NXF, QH, DJ and SHN acknowledge the financial support by the NSF (Grant No. CMMI-1120724) and AFOSR MURI (Award No. FA9550-12-1-0488). NXF and SHN are also supported by the Cooperative Agreement between the Masdar Institute of Science and Technology (Masdar Institute), Abu Dhabi, UAE and the Massachusetts Institute of Technology (MIT), Cambridge, MA, USA -  Reference 2/MI/MI/CP/11/07633/GEN/G/00. XZ and JX acknowledge the financial support from AFOSR MURI (Award No. FA9550-12-1-0488). YL acknowledges the support of 3M Non-Tenured Faculty Award. QH would like to thank Tresback Jason in the Center for Nanoscale Systems (CNS) for his help on AFM characterization.

% Bibliography
\bibliography{Reference}

\end{document}